\documentclass[prl,twocolumn,showpacs,superscriptaddress]{revtex4-1}
\usepackage{graphicx,amssymb,amsmath,psfrag,color}
\usepackage[colorlinks=true,citecolor=blue,linkcolor=blue,urlcolor=blue]{hyperref}

\newcommand{\la}{\langle}
\newcommand{\ra}{\rangle}
\newcommand{\lb}{\left|}
\newcommand{\rb}{\right|}

\newcommand{\sz}{S^z}
\newcommand{\sx}{S^x}
\newcommand{\sy}{S^y}

\begin{document}

\title{How periodic driving heats a  disordered quantum spin chain}
\author{Jorge Rehn, Achilleas Lazarides, Frank Pollmann, Roderich Moessner}
\affiliation{Max-Planck-Institut f\"ur Physik komplexer Systeme, 01187 Dresden, Germany}
\date{\today}

\begin{abstract}
We study the energy absorption in real time of a disordered quantum spin chain subjected to
coherent monochromatic periodic driving. We determine characteristic
fingerprints of the 
well-known ergodic (Floquet-ETH for slow driving/weak disorder) and many-body localized
(Floquet-MBL for fast driving/strong disorder) phases.  In addition, we identify 
an intermediate regime, where the energy density
of the system -- unlike the entanglement entropy a local and bounded observable -- grows
{\it logarithmically} slowly over a very large time window.
\end{abstract}
\maketitle

{\bf Introduction.} Periodically-driven, or Floquet, many-body quantum systems are a current focus of out-of-equilibrium physics. Generically, an external forcing pushes the system away from equilibrium and heats up the system, as is natural for a non-adiabatic perturbation.  Ergodic many-body systems in particular heat up to reach a fully-mixed state (also known as  ``infinite-temperature'' or Floquet-ETH state)~\cite{Lazarides:2014ie,DAlessio:2014fg,Ponte:2015hm}. Interestingly, such a heat death of the correlations can be avoided in the presence of constraints frustrating the entropy increase. One  possibility is in a Floquet-integrable system where there exist quantities  conserved even in the presence of driving. This leads to a synchronised state  maximizing entropy, but now subject to constraints imposed by those conserved quantities~\cite{Lazarides:2014cl}; this is known as periodic Gibbs ensemble, in analogy
to the generalised Gibbs ensemble of static systems~\cite{RigolGGE}.

An alternative, more robust way to prevent full heating -- not requiring  the fine-tuning 
needed for integrable or dynamically localized~\cite{Das:2010dy,Eckardt:2009gz,HaenggiReview} behavior -- is provided by
many-body localization (MBL) \cite{Basko:2006hh, Gornyi:2005,Oganesyan:2007ex}. 
Here, the addition of  sufficiently strong disorder to
an interacting, ergodic system leads to vanishing energy, particle and
spin transport so that ergodicity is broken in the static (time-independent)
limit. When switching on periodic driving, a region
in the driving frequency-amplitude plane exists in which  the system approaches a state that is not fully mixed, and in particular has finite energy with respect to, e.g., the static, or average, Hamiltonian~\cite{Lazarides:2015jd,Ponte:2015dc}. In this regime the effective Hamiltonian governing the stroboscopic dynamics 
may exhibit sharply distinct phases, characterized by order parameters, including ones  with no equilibrium counterparts~\cite{Khemani:2015wn}.

\begin{figure}
  \centering
	\includegraphics[scale=0.8]{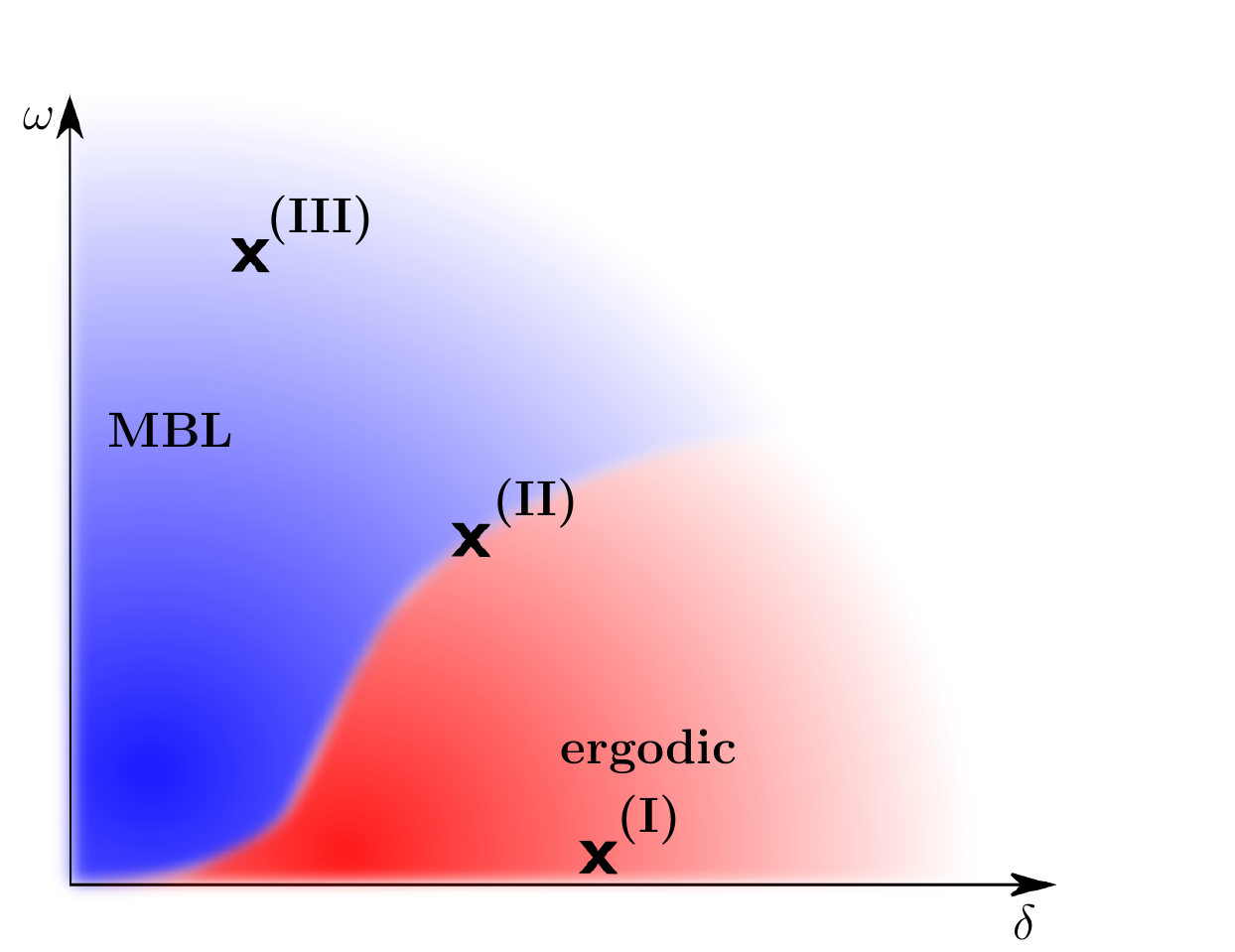}
	\caption{(color online) Proposed phase diagram for
the long time state of a driven strongly, disordered system as a function of
the driving frequency $\omega$ and strength $\delta$. Red (I)
and blue (III) indicate Floquet-ETH and
Floquet-MBL behavior, where the system approaches a fully-mixed,
``infinite-temperature'' state or remains localized, respectively.
At (II), heating leads to energy growing
logarithmically slowly with time (Fig.~\ref{fig:low-highfreq}(II)) over a broad time window.
Our numerics cannot access small $\omega$
and $\delta$ near the origin where Ref.~\onlinecite{Abanin:2014te}
suggests the indicated behavior.
	}
	\label{fig:diagram}
\end{figure}

What this classification leaves open entirely 
is how the process of synchronization takes place, i.e.\ how  the steady state is approached in real time. 
This question is not only of intrinsic fundamental importance, but it also occurs in the context of practical applications,
such as in Floquet-engineering cold atomic systems \cite{Eckardt:2008fg,Holthaus2015a}, where one is interested to what extent
generating interesting effective Hamiltonians unavoidably goes along with heating \cite{Genske:2015iq}.

In this article, we analyze this process in a setting which combines disorder, interactions, and driving. In particular, we study how energy is absorbed in real time. We consider  a disordered spin chain initially in the ground state of a static Hamiltonian, and monitor stroboscopically  its energy density with respect to this Hamiltonian upon switching on the periodic drive. 

In the {strongly disordered} case, with parameters chosen such that the static model is MBL, we confirm the existence of the localized and ergodic regimes under driving, and describe characteristic properties of their heating process.  In the ergodic regime, we find that the energy indeed saturates at the maximum-entropy, ``infinite-temperature'' value (as it does when starting from the weakly disordered ergodic regime); the rate at which this is reached scales quadratically with the driving amplitude, $\delta$, consistent with a Fermi Golden rule-type picture.  

In the localized regime driving results in a non-zero excess energy density, $\epsilon_\infty$, which is 
reached relatively swiftly. For a driving frequency, $\omega$, well above the otherwise dominant local disorder
strength, $\eta$, we find an asymptotic dependence 
$\epsilon_\infty\propto\delta^2/\eta^2$, varying parameters beyond which leads to deviations
presaging the delocalization transition. This can be understood via the behaviour of driven two-level
systems. 

Most remarkably, at the crossover between the two we 
find a {\it logarithmically slow} heating process, with energy entering the system over a window 
extending over several decades in time. This is superficially 
reminiscent of the characteristic logarithmic growth of the 
entanglement entropy 
in a static MBL system \cite{Znidaric:2008ux,Bardarson:2012gc,Serbyn:2013he,Vosk2013,Nanduri2014}. 
However, the internal energy is a quantity which is a local observable, for which no such 
result is known. Also, given the boundedness of the internal energy in the model we study, the 
time window of the logarithmic growth is necessarily limited; however, this phenomenon is 
comfortably visible in our numerics already for chains of length $L=20$; 
an extrapolation of our results suggests it may extend over a window
extending over more than eight decades in time. 

\begin{figure*}[t!]
	\includegraphics[scale=0.46]{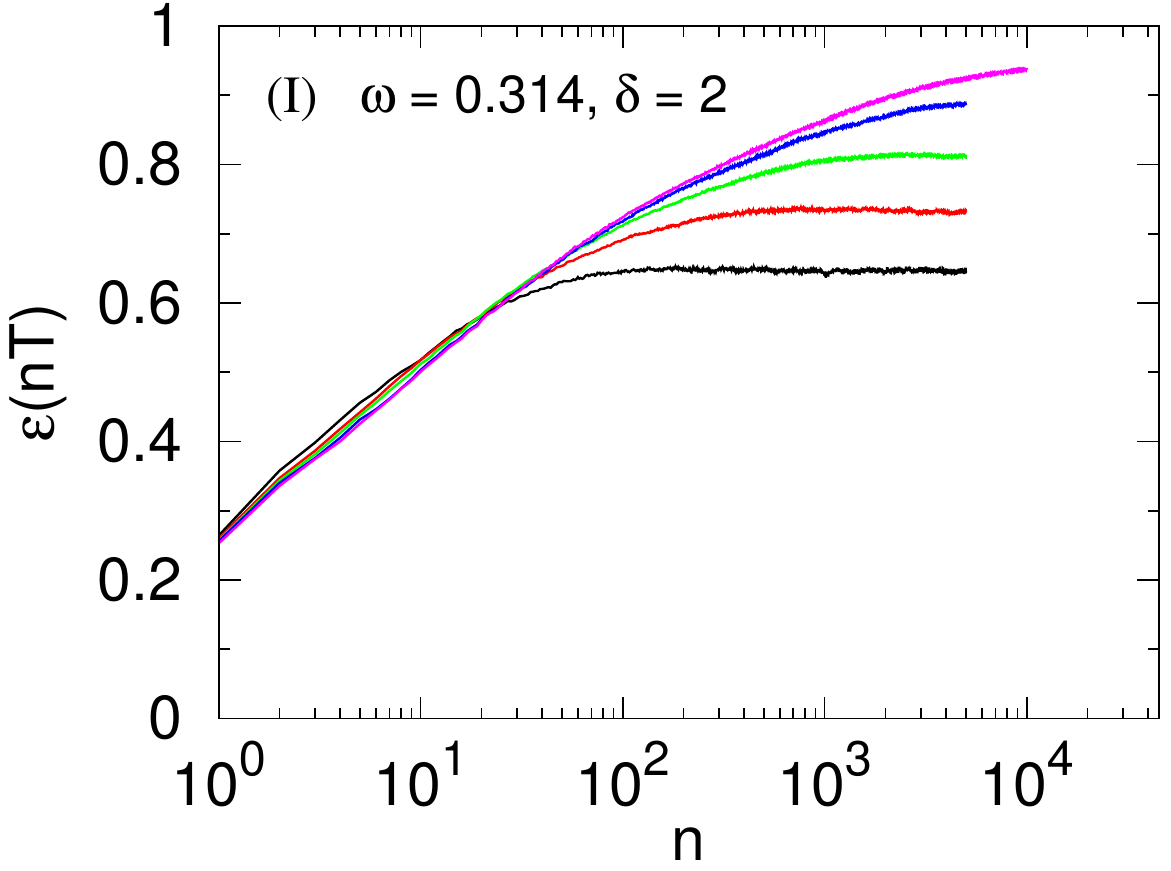}
	\includegraphics[scale=0.46]{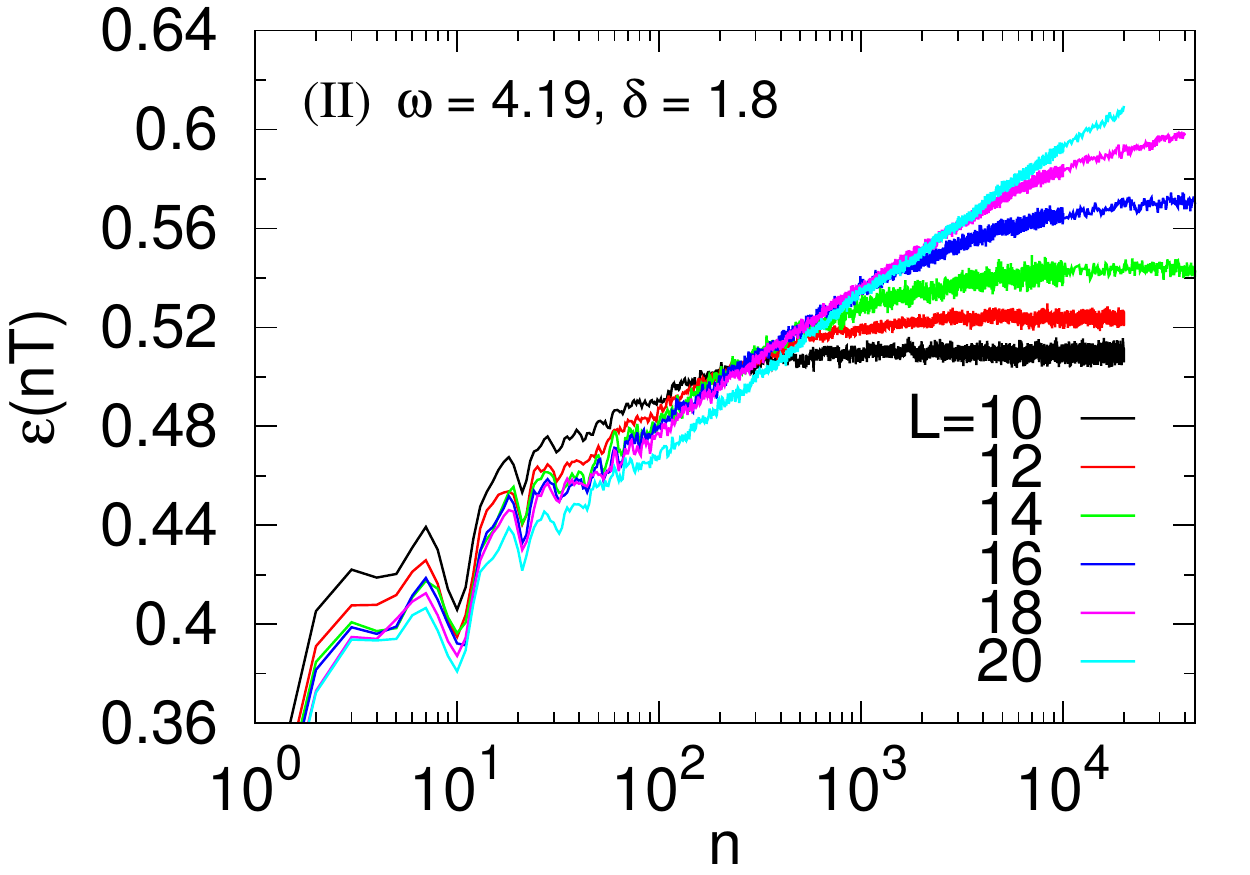}
	\includegraphics[scale=0.46]{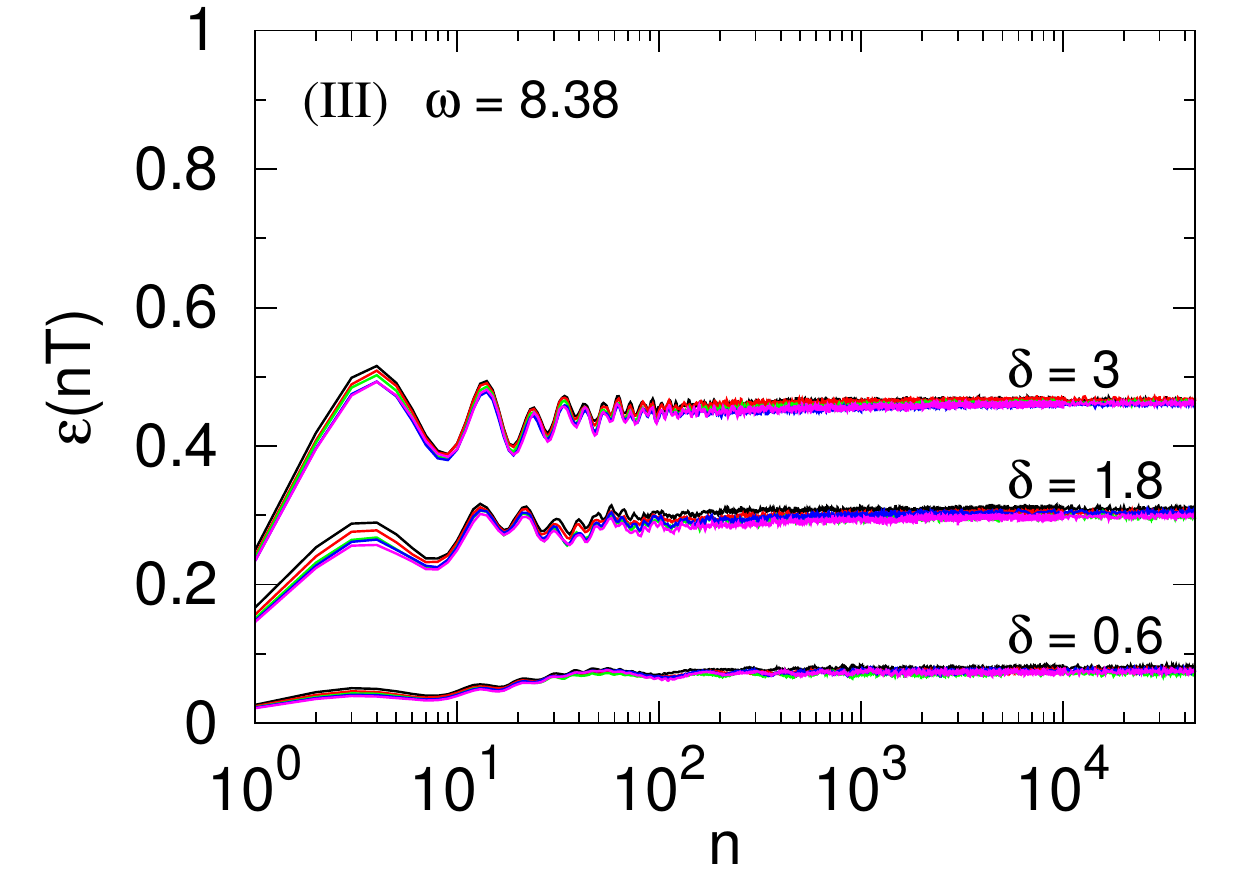}
	\caption{(color online).
	Energy absorption in a strongly disordered system with $\eta=5$ and $J_z=0.5$, corresponding to 
	different locations in the phase diagram, Fig.~\ref{fig:diagram}. At (I), an initially MBL system
delocalizes and heats up to a fully-mixed state.
	In the intermediate regime, (II), the system
heats up to the fully-mixed state, but logarithmically slowly. This
slow growth persists for longer times as we increase $L$.
	For Floquet-MBL, (III), driving does not
delocalize the system, leading instead to a localized long-time state which
has partially heated up to some intermediate energy.
Fig.~\ref{fig:epsvsDeltaLoc} explores the dependence of the final energy
on the driving amplitude $\delta$.
	}
	\label{fig:low-highfreq}
\end{figure*}

{\bf Model.} We study the spin-1/2 XXZ chain in a disordered longitudinal field
subject to a monochromatically driven staggered field with period $T=2\pi/\omega$:
\begin{eqnarray}
H(t) & = & H_{0} + H_D(t)   \label{eq:ham_drive} \\
	H_{0} &=& J_{\perp}\sum_{i=0}^L (S^x_iS^x_{i+1} + S^y_iS^y_{i+1}) + J_z\sum_{i=0}^L S^z_iS^z_{i+1}\\
	& & +  \sum_{i=0}^L h^z_iS^z_i,
\end{eqnarray}
where $h^z_i\in[-\eta,\eta]$, $J_{\perp},J_z\ge0$,
and with driving
\begin{align}
H_{D}(t)= -\delta\cos{\omega t}\sum_{i=0}^L (-1)^i S^z_i .
  \label{eq:ham_dri}
\end{align}

The static part $H_0$ is known to be MBL  for $J_z\neq0$
and sufficiently strong disorder $\eta>\eta_c$~\cite{Oganesyan:2007ex,Pal:2010gr}.
Driving such MBL systems is expected to 
lead to delocalization ~\cite{Lazarides:2015jd,Ponte:2015dc} for low enough frequencies at fixed $\delta$,
yielding the phase diagram schematically shown in
Fig.~\ref{fig:diagram}, with a high-frequency regime Floquet-MBL and a low-frequency 
Floquet-ETH regime where a fully-mixed state is approached \cite{Abanin:2014te}.

\begin{figure}[t]
	\includegraphics[scale=0.55]{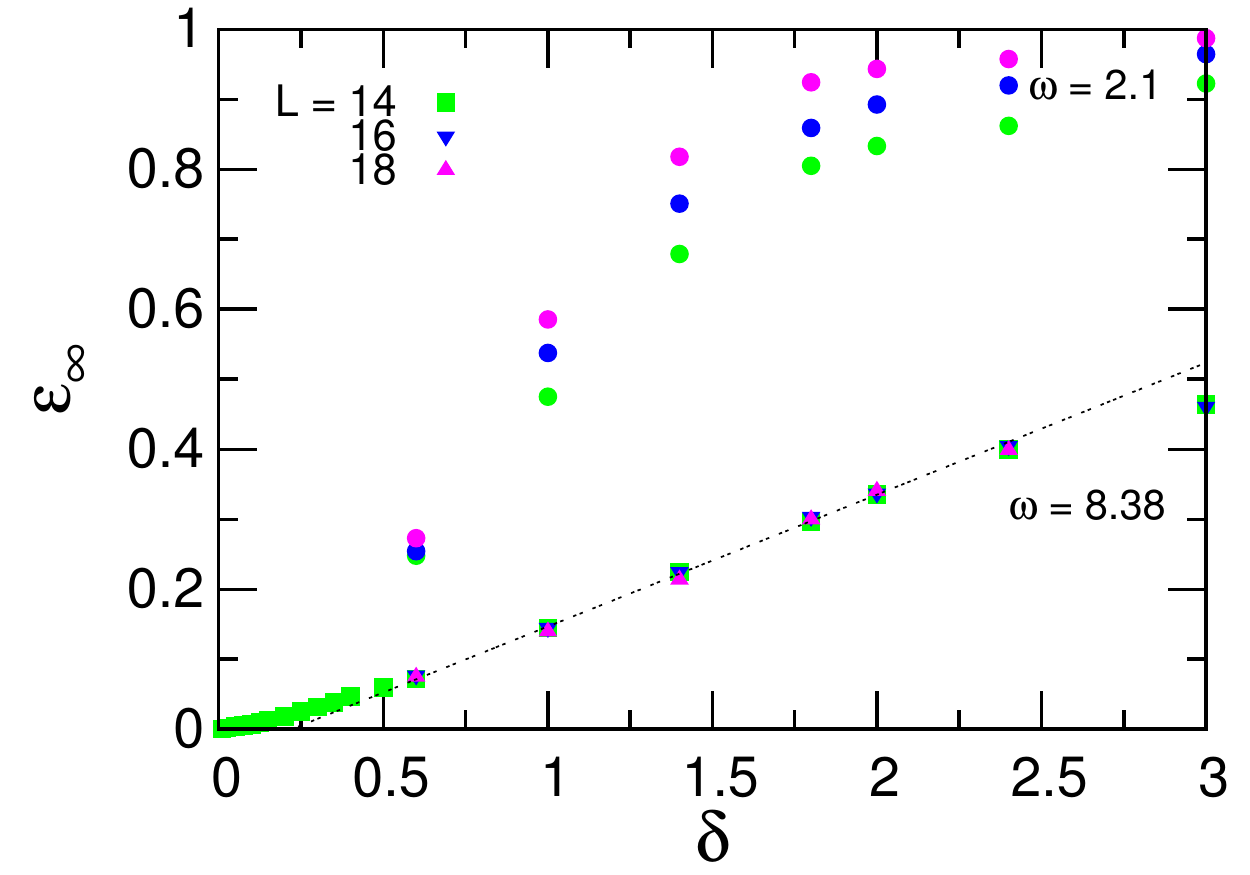}
	\includegraphics[scale=0.55]{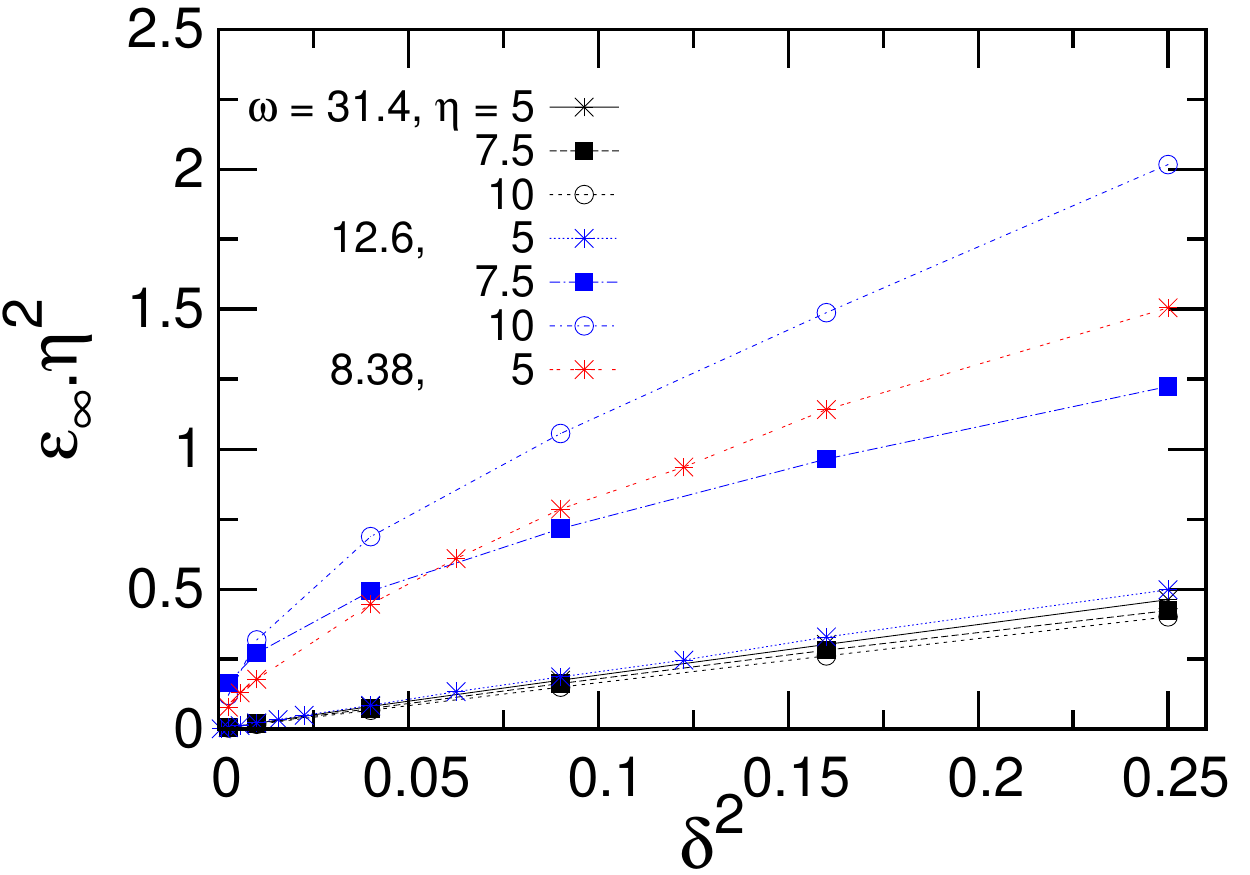}
	\caption{(color online). Upper panel: Final excess energy density
$\epsilon_\infty$ vs. driving amplitude $\delta$ for two values of the
frequency, interaction strength $J_z=0.5$ and disorder strength $\eta=5$
corresponding to MBL in the static case. For the upper set of data,
corresponding to the  Floquet-ETH regime, increasing system size
leads to increasing $\epsilon_\infty$, so that 
it approaches 1 in the thermodynamic limit, the fully-mixed result. For the lower group of curves,
corresponding to the localized regime, the energy saturates to a finite value $\epsilon_\infty<1$
 depending
on $\delta$, implying a limited absorption of energy and the presence of a finite localization
length. 
Lower panel: Same quantity for large values of $\omega$, in a
regime of small driving amplitudes. For frequency sufficiently large compared to $\eta$  (lower set of points) scaling collapse of data corresponding to different disorder amplitudes $\eta$ is possible ($L=14$). In this panel, points of the same color have the same $\omega$ while points of the same shapes have the same $\eta$.
}
	\label{fig:epsvsDeltaLoc}
\end{figure}

Here we concentrate on real-time dynamics, in particular the
(stroboscopic) time evolution of the energy in the system,
expressed via the rescaled excess energy density:
\begin{align}
	\epsilon(nT) = \frac{\la\psi\lb H(nT)\rb\psi\ra-E_{\mathrm{min}}}{\overline{E}-E_{\mathrm{min}}},
\end{align}
with $\overline{E}=D_H^{-1}\mathrm{tr}[H(0)]$ and $D_H$ being the
 Hilbert space dimension, so that $\epsilon=0$ in the ground
state of $H(0)$, while $\epsilon=1$ for the fully mixed 
ensemble. 

In what follows
we discuss in detail different parameter regimes and characterize their dynamical properties. 
In the case of strong disorder, where the static system
is in the MBL regime, we study three points on the phase diagram in Fig.~\ref{fig:diagram}, 
corresponding to the above mentioned behaviors (I), (II) and (III), 
in detail numerically by finite-size simulations.

For all our numerical studies we initialize
the system in the ground state of the static Hamiltonian at
$t=0$, $H(0)$, so that $\epsilon(0)=0$ and fix the interaction
strength to $J_{\perp}=1$ and $J_z=0.5$.
The ground state is obtained from the sparse matrix representation of the static Hamiltonian using the Lanczos method and for the time evolution of the driven system we use an iterative Krylov space based algorithm \cite{saad:1992}.  The harmonic driving is discretized using sufficiently small time steps of $\delta t = 0.0250 J_{\perp}^{-1}$ ($\delta t = 0.0025 J_{\perp}^{-1}$ for large frequencies). We  average the results over $N_{\mathrm{disorder}}=300$-$500$ disorder realizations. For all simulations we utilise the conservation of $S_z^{\mathrm{total}}$ and consider only the $S_z^{\mathrm{total}}=0$ sector, allowing for a considerable speedup.

{\bf Strong disorder.}
We begin by setting the disorder strength to $\eta=5$, which puts the static Hamiltonian comfortably in its MBL phase, and 
choose a driving amplitude $\delta=2$.
For a driving frequency of $\omega=0.314$, corresponding to (I) in Fig.~\ref{fig:diagram}, we find that the system is in the
ergodic regime, approaching the ``infinite
temperature'' state, with energy $\overline{E}$ corresponding to $\epsilon_\infty=1$, in the long time limit 
(Fig.~\ref{fig:low-highfreq}, leftmost panel).
 While finite-size effects are visible at the sizes displayed,
there is  convergence to   $\epsilon_\infty=1$ with increasing system size. 

Next we increase the frequency to $\omega=8.38$ and choose
$\delta=0.6, 1.8, 3.0$, finding that all these points lie in
the Floquet-MBL phase corresponding to the neighborhood of point
(III) in Fig.~\ref{fig:diagram}. Here the system at first absorbs
energy comparatively quickly but stops well short of the 
``infinite-temperature'' point (Fig.~\ref{fig:low-highfreq}, rightmost panel). 
The saturated value of $\epsilon_\infty$ depends on
the system parameters, and increases with $\delta$. This is the
Floquet-MBL phase and it is stable for a range of $\delta$.
Notice also that there is very little system-size dependence in
the results, indicating that for the Floquet-MBL regime finite-size
effects are much weaker than for the ergodic regime of panel (I), as 
expected for a system with a finite localization length.
This is also consistent with the results of Ref.~\cite{Lazarides:2015jd},
where level statistics results also lie in-between the localized
(Poisson) and ergodic (circular unitary ensemble) results. In
this regime ergodicity is broken, there exist local integrals of
motion and therefore the final state is not fully-mixed. We   
explore the  dependence of the absorbed energy on
$\delta$ further down.

Finally we select $\omega=4.19$ and $\delta=1.8$, corresponding to
point (II) in Fig.~\ref{fig:diagram} which lies in between the
well-localized and comfortably ergodic regimes. While the energy density grows to its
``infinite-temperature'' value as in the ergodic case (I), its
logarithmically slow growth extends over several decades
(Fig.~\ref{fig:low-highfreq}, middle panel)! 
This logarithmic growth strikingly visible in the crossover regime has not been observed before and is the
central result of our paper. 
We emphasize that this phenomenon is only superficially reminiscent of the  logarithmic growth of the entanglement entropy  described in
Ref.~\cite{Bardarson:2012gc,Znidaric:2008ux} and later on also observed in Floquet-MBL systems \cite{Ponte:2015dc}. The entanglement growth is
by now well-understood in terms of the structure of the 
MBL Hamiltonian in terms of local ``l-bits''~\cite{Huse:2014co,Serbyn:2013ug,Chandran:2015cw,Ros:2015ib}.
Most fundamentally, the quantity $\epsilon_\infty$ we consider is a fully local 
observable. Further, it is bounded, so that the logarithmic growth
eventually has to terminate. However, this leaves a very large
scope for the logarithmic growth---extrapolating the growth
in the middle panel of Fig.~\ref{fig:low-highfreq} all the way until  
the maximum 
$\epsilon_\infty=1$ is reached allows for 
a time window covering 8 or 9 decades. 

The slow logarithmic growth in regime (II)
is astonishingly pronounced.   By contrast, the 
curves for (I) and particularly (III) plateau out much earlier, so that
it is not possible to determine whether the rise is sensibly
described by a quickly-terminated logarithmic growth akin to the slowly-terminated window in (II),
or rather by a different functional form.
We do note that the curves for all values of $L$ agree
for short times in Fig.~\ref{fig:low-highfreq}(I), defining a limiting curve in the thermodynamic limit
from which finite-size systems peel off
at a time which grows with $L$. The shape of the limiting curve is not inconsistent
with logarithmic growth, albeit over an inconclusively limited timespan.

Let us now turn to  the
behavior of $\epsilon_\infty$ as a function of $\delta$, plotted
in Fig.~\ref{fig:epsvsDeltaLoc} for different system sizes $L$
(upper panel). 
In the ergodic regime (upper set of points) and for sufficiently
large $\delta$, the energy density $\epsilon_\infty$ increases with increasing system
size $L$, indicating that it will approach the fully-mixed value $\epsilon_\infty=1$
for large enough $L$.

For smaller $\delta$, $\epsilon_\infty$ still increases with
$L$ but more slowly. This is consistent with the phase diagram in
Fig.~\ref{fig:diagram}, since a smaller $\delta$ means the system is closer
to the ergodic regime. The finite-size numerics reported here
cannot determine whether or not there eventually is a critical $\delta$ below
which MBL survives for any value of driving frequency $\omega$.

In the localized regime (lower set of points) the saturation value
grows approximately linearly for an intermediate range of $\delta$, sandwiched 
between ergodic saturation for large $\delta$, and a quadratic regime for small $\delta$. 
Here, the saturation value does not depend
on system size, indicating the presence of an appropriately defined localization length beyond
which an increase in $L$ no longer changes $\epsilon_\infty$, so that  
even in the thermodynamic limit
the system absorbs only a finite amount of energy per unit length but
stops short of heating up to completely. 
This behavior
is consistent with the presence of local integrals of motion~\cite{Imbrie:2014vo,Serbyn:2013ug,Huse:2014co,Chandran:2015cw,Ros:2015ib} in the
effective Hamiltonian. These can then play the role of 
conserved quantities for the driven
problem, restricting in turn the growth
of entropy and thus making the fully-mixed state with $\epsilon_\infty=1$ inaccessible for the given initial condition. The accessible region of Fock space grows as the transition point is approached, leading to the observed increase of $\epsilon_\infty$.

The lower panel of Fig.~\ref{fig:epsvsDeltaLoc} shows the
final value of the energy density for large values of $\omega$ focusing
on the region of asymptotically small driving amplitudes. For frequencies large
enough (but still below the many-body bandwidth of our finite-size systems), data corresponding to
different disorder strengths $\eta$ collapse and scale  
with $\delta^2/\eta^2$ (lower set of points). This dependence is in agreement with that exhibited by a set of independent, 
driven two-level systems, again indicating that the local integrals of motion do indeed survive in the Floquet-MBL system,
as described in the appendix. Once this behavior
sets in, transitions to higher energy states are suppressed, and the system
absorbs almost no energy. 

\begin{figure}[t!]
	\includegraphics[scale=0.55]{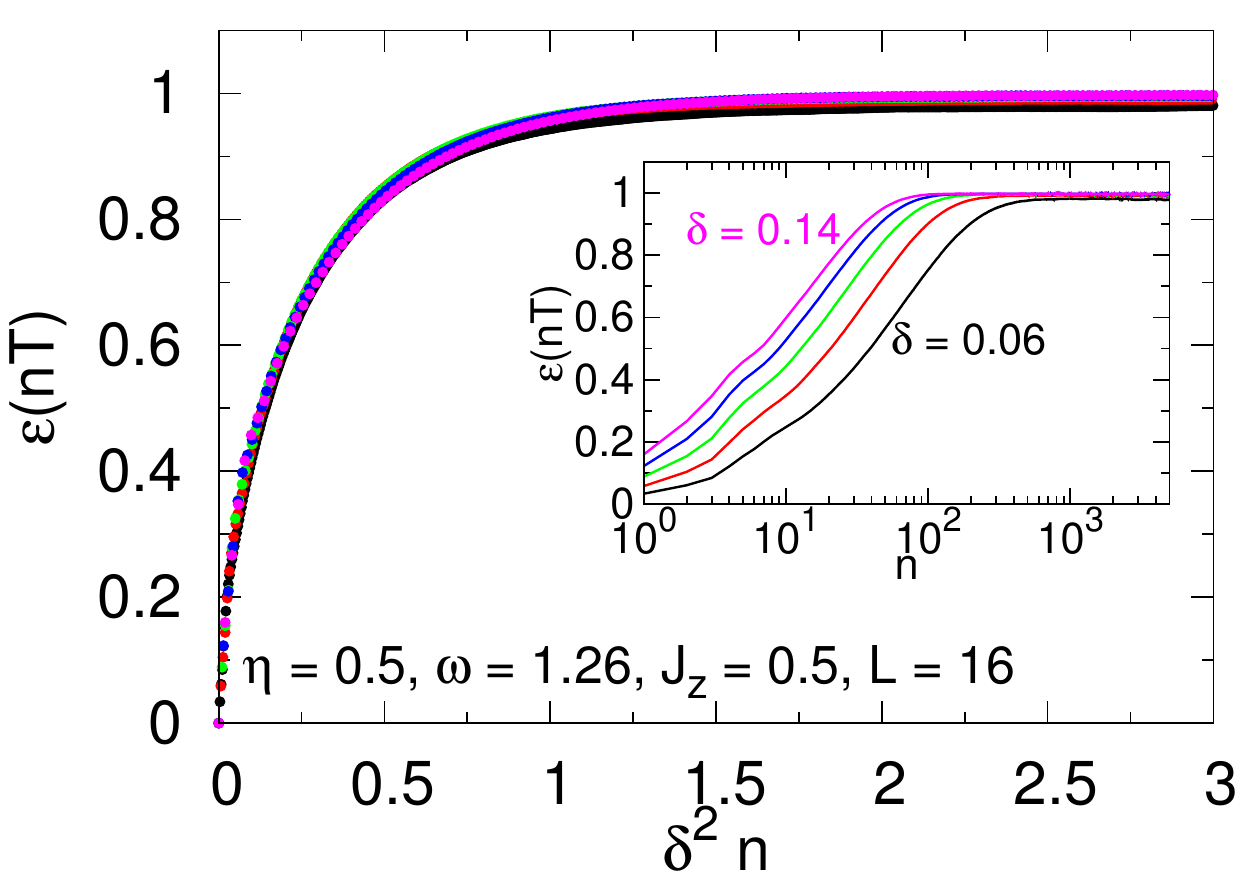}
	\caption{(color online).
		Energy absorption for different driving amplitudes in a weakly disordered system with $\eta=0.5$ and $J_z=0.5$. The inset
shows the bare data while the main plot shows the collapsed data after
rescaling the time coordinate by $\delta^2$.
	}
	\label{fig:ergodicregime}
\end{figure}

{\bf Weak disorder.} 
We finally turn to the case where the static Hamiltonian itself is still
disordered but not in the MBL phase, Fig.~\ref{fig:ergodicregime}.
In this regime the system fully heats up as in the clean
case~\cite{Lazarides:2015jd}. The inset in Fig.~\ref{fig:ergodicregime}
shows the bare data while the main plot has the time
rescaled by a factor of $\delta^2$. The heating thus depends on the 
amplitude $\delta$ and time $n$ (measured
in units of the stroboscopic step) via the combination
$\delta^2 n$. This is consistent with the
expectation from leading-order perturbation theory (as in Fermi's Golden Rule),
which gives the observed dependence on the square of the driving amplitude
$\delta$. This is in contrast to the Floquet-MBL case (Fig.~\ref{fig:low-highfreq}) where  no such 
collapse occurs, demonstrating the
breakdown of linear response for the long-time behavior of the system.


{\bf Conclusions}
We have studied the energy absorption in real time of a disordered quantum spin chain subjected to
coherent monochromatic periodic driving in different parameter regimes. 
For strongly disordered systems, in which the static Hamiltonian is in the many-body localized phase, we have identified three regimes: An ergodic regime  in which the system  heats up to ``infinite temperatures''; a well-localized regime in which the system quickly plateaus at  some finite  energy density; and an intermediate  regime in which the system slowly heats up with a logarithmic increase of energy over several decades. This logarithmic growth is very distinct from the characteristic logarithmic growth of entanglement entropy in that the energy density is a locally observable quantity.
For weakly disordered systems,  where the static Hamiltonian is in the extended phase, we observe that energy is quickly absorbed until reaching a fully mixed state, with the heating curves collapsing upon rescaling time with a factor $\delta^2$. Also, in the strongly disordered case, driving 
at high frequencies yields to behaviour which can be understood in terms of driven two-level systems. 

Our results provide the first detailed study of the energy absorption over time in a Floquet-MBL system. It ties in with the broader interest in how Floquet systems reach their steady states, where an increasingly rich phenomenology is being uncovered. This also comprises the case  of {clean systems and fast driving, where it has been argued that approaching  the fully mixed state can be extremely slow~\cite{Abanin:2015bc,Mori:2015wb}. It is an open question whether the pronounced logarithmic growth we have uncovered might be related to the glassy behavior seen in Ref.~\cite{Bukov:2015} for a clean system, where the authors argue that the appearance of rare resonances which eventually proliferate are what causes the heating. The extent to which this set of phenomenologies generalizes to higher dimension remains a tantalizing open question, with our capacity to find an answer limited by the usual difficulties in treating systems combining interactions and disorder. 

More broadly, our work advances our understanding of the phenomenon of MBL as well as of the properties of the newly 
discovered Floquet ensembles, both of which continue to constitute intensely studied and rapidly advancing subfields of out-of-equilibrium many-body quantum dynamics.

{\bf Acknowledgements.} This work was in part supported by DFG via SFB 1143. A. L. and R.M. 
are grateful to Arnab Das for collaboration on related work, and to Vedika Khemani and
Shivaji Sondhi for useful discussions.  

{\bf Note added.} As we were writing up this work, a preprint \cite{1602.06055} appeared which
studies the question of energy absorption in an MBL system driven by a strong field. At the same time as this work, a preprint
is being prepared which focuses on heating in the Floquet-MBL phase \cite{Gopalakrishnan:2016}.

{\bf Appendix.} Here, we provide details of the computation of $\epsilon_\infty\sim\delta^2/\eta^2$ for a driven two-level system
 to model the strongly disordered regime at high frequency.
The driven two-level system is described by
\begin{equation*}
        H(t) = \frac{1}{2}\eta \sz + \delta\cos(\omega t)\sx.
\end{equation*}
(note that $\eta$ here is related, but not identical, to the disorder 
amplitude in the main text). In the high-frequency regime, 
$\omega\gg\eta\gg\delta$, we calculate the time-averaged energy of this 
system with the ground state of the $t=0$ Hamiltonian as the initial state. 
To this end we find the Floquet Hamiltonian $H_F$ to leading order in 
the Magnus expansion~\cite{Eckardt:2015hp,Bukov:2014gu},
\begin{equation*}
        H_F = h_z \sz+h_x\sx.
\end{equation*}
with
\begin{eqnarray}
 h_z&=&\frac{1}{2}\eta + \frac{1}{8} \frac{\delta^2\eta}{\omega^2}\\
h_x&=&\frac{\delta\eta^2}{2\omega^2}.
\end{eqnarray} 
This may be diagonalised as follows:
 a rotation $U=\exp\left(i \sy \theta\right)$ with 
$\theta=\arctan(h_x/h_z)$ aligns $H_F$ with the $z$-spin axis, whence 
the eigenvectors and eigenvalues may be read off. Rotating back, the 
eigenvectors and eigenvalues are $| e_{\uparrow/\downarrow}\ra = 
U^\dagger | \uparrow/\downarrow \ra$, with $| \uparrow/\downarrow \ra$ 
the eigenvectors of $\sigma^z$, and $\pm \sqrt{h_x^2 + h_z^2}$, respectively.sf 

The initial state is the ground state of 
$H(0)=\frac{1}{2}\eta \sz + \delta\sx$, which can be written as 
$|\psi_0\ra = V^\dagger |\downarrow\ra$ with $V = \exp\left(i\sy 
\arctan(2\delta/\eta)\right)$. The time-averaged energy starting from 
this initial state is given by
\begin{equation*}
        E_\infty = \sum_{\alpha=\uparrow/\downarrow}
                                        \la e_\alpha | H(0) | e_\alpha \ra
                                                \left| \la \psi_0 | e_\alpha \ra \right|^2        
\end{equation*}
Rescaling this as in the main text and expanding to leading order in the 
parameters $\delta/\omega$ and $\delta/\eta$ we obtain
\begin{equation*}
        \varepsilon_\infty = 2\frac{\delta^2}{\eta^2}
\end{equation*}
which has the same form as the high-frequency (i.e., the lowest set of) curves of the lower 
panel of Fig. 3 of the main text.

\bibliographystyle{apsrev4-1}
\bibliography{engrowth}
\end{document}